\documentclass[%
 aip,
 amsmath,amssymb,floatfix,
 reprint,%
]{revtex4-2}

\usepackage{graphicx}
\usepackage{dcolumn}
\usepackage{bm}

\usepackage[utf8]{inputenc}
\usepackage[T1]{fontenc}
\usepackage{mathptmx}
\usepackage{etoolbox,xcolor,booktabs}
\usepackage{multirow}
\makeatletter
\patchcmd{\titleblock@produce}
  {\frontmatter@RRAPformat}
  {\frontmatter@RRAPformat{\produce@RRAP{*Corresponding author:\ \href{mailto:sajid.ali@uq.edu.au}{sajid.ali@uq.edu.au}}}\frontmatter@RRAPformat}
  {}{}
\makeatother

\begin{document}

\preprint{AIP/123-QED}

\title[A Flexible Kinetic Monte Carlo Framework for GaN Molecular Beam Epitaxy with Adaptive On-the-Fly Barrier Evaluation]
{A Flexible Kinetic Monte Carlo Framework for GaN Molecular Beam Epitaxy with Adaptive On-the-Fly Barrier Evaluation}

\author{Sajid Ali*}
\affiliation{School of Mathematics and Physics, The University of Queensland, Brisbane 4072 Queensland, Australia}
\email{sajid.ali@uq.edu.au}

\author{Norbert Krause}
\affiliation{Silanna Semiconductor Pty Ltd, Pinkenba 4008 Queensland, Australia}

\author{Carla Verdi}
\affiliation{School of Mathematics and Physics, The University of Queensland, Brisbane 4072 Queensland, Australia}

\begin{abstract}

We present a lattice-based kinetic Monte Carlo (KMC) framework for simulating GaN(0001) growth by molecular beam epitaxy. The framework captures the key microscopic processes governing epitaxial growth, including temperature-dependent surface diffusion, flux-driven deposition, Ehrlich--Schwoebel (ES) step-edge barriers, Ostwald ripening, and species-specific desorption, within a scalable 
architecture that enables systematic exploration of experimentally relevant growth conditions. 
In addition to predefined activation-energy catalogs, the framework supports adaptive on-the-fly barrier evaluation using machine-learned interatomic potentials. When previously unencountered local atomic configurations arise, activation barriers are computed via nudged elastic band, potential energy scans, or Br{\o}nsted--Evans--Polanyi methods, and cached for reuse. 
Predefined-barrier simulations reproduce compact triangular island formation, and further capture Ostwald ripening during growth interruptions and ES barrier-induced multilayer nucleation. At elevated temperatures, desorption drives an island ``walking'' regime, in which N--Ga exchange generates weakly bound Ga adatoms (AdGa) at trailing edges; preferential desorption of AdGa leads to asymmetric edge retreat and net island translation. Our KMC framework provides a flexible platform for predictive simulations of GaN epitaxy at the atomic scale and, more broadly, non-equilibrium growth of compound semiconductors.

\end{abstract}

\maketitle

\section{Introduction}

Surface morphology during compound semiconductor epitaxy is governed by the interplay between multi-species diffusion, attachment asymmetries, and temperature-dependent kinetics. In GaN molecular beam epitaxy (MBE), island formation, step-flow evolution, coarsening, and desorption-driven mass loss all arise from processes that operate far from equilibrium and span disparate length and time scales. Atomistically informed kinetic modeling is therefore essential for mechanistic insight into these phenomena. This is particularly important for GaN and AlGaN-based heterostructures used in deep-UV optoelectronics\cite{moustakas2017optoelectronic}, where surface kinetics during MBE directly control interface abruptness and defect incorporation in short-period superlattices\cite{nicholls2023high}, ultimately governing optical efficiency and emission wavelength\cite{piva2020modeling}.

Kinetic Monte Carlo (KMC) methods are widely used to simulate epitaxial growth, describing surface evolution as a sequence of stochastic, thermally activated events whose rates are determined by activation barriers \cite{Voter2007,Fichthorn1991, evans2006morphological}. In compound systems such as GaN, however, growth morphology depends sensitively on several processes. These include coordination-dependent diffusion pathways, asymmetries in interlayer mass transport, and interactions between multiple atomic species. A key example is the Ehrlich--Schwoebel (ES) barrier, which hinders downward step crossing and thereby induces asymmetric interlayer mass transport, destabilizing step-flow growth and promoting mound formation \cite{kaufmann2016critical}. At elevated temperatures, desorption competes with the incoming flux, modifying both the effective growth rate and surface stoichiometry \cite{bruno2010adsorption, lymperakis2024desorption}. Over longer timescales, 
mass transport from smaller to larger islands leads to surface coarsening through Ostwald ripening \cite{voorhees1985theory}. Although these effects have been studied individually, a unified computational framework capable of treating them simultaneously is still lacking.

Traditional lattice KMC approaches rely on predefined catalogs of activation barriers derived from density functional theory (DFT) calculations or empirical parameterizations~\cite{chugh2017lattice, ranganathan2025ga}. 
While computationally efficient, such static catalogs restrict diffusion to predefined coordination classes and cannot systematically account for the diversity of local environments that arise during realistic growth, particularly at step edges
and evolving island boundaries. On-the-fly (OTF) evalua-
tion of activation barriers, where transition energies are computed dynamically from the instantaneous local atomic environment, offers a route to overcome these limitations~\cite{yokaichiya2024fly}.

Several related approaches have been developed in the context of compound semiconductors. 
Thomas~\textit{et al.}\cite{thomas2024multiscale} 
presented a comprehensive DFT+KMC multiscale model coupling NEB-derived barriers with a gas-phase transport model to describe Ge and oxygen incorporation on GaN. 
Ferreyra and Quiroga\cite{ferreyra2021} developed the ``Assistant KMC'' framework, a parametric KMC tool for Ge/Ga/N co-deposition on GaN(0001) that employs configuration-dependent Arrhenius rates. 
Thomas~\textit{et al.}\cite{thomas2025twobody} subsequently introduced a minimum direct two-body barrier approach spanning both surface and bulk processes. The present work is complementary to these impurity-focused studies, as it targets native-species (Ga and N) surface morphological evolution and extends barrier evaluation on the fly to novel coordination environments encountered at step edges\cite{yokaichiya2024fly}. The self-learning KMC method proposed by Trushin~\textit{et al.}\cite{trushin2005slkmc} pioneered on-the-fly barrier caching for metallic surfaces, combining binary-encoded local-environment fingerprints with explicit saddle-point searches to build a self-updating barrier database. Building on this concept, our framework 
extends the paradigm to a three-species compound semiconductor using a species-resolved coordination fingerprint and accelerated OTF barrier evaluation based on machine-learned interatomic potentials.

In this work, we develop a lattice KMC framework for GaN(0001) growth that integrates terrace diffusion, ES barriers, species-dependent desorption, and long-time coarsening within a single simulation environment. Using a coordination-based predefined barrier catalog, we first simulate temperature-driven morphology evolution, ES-barrier-induced multilayer growth, desorption-driven edge roughening, and Ostwald ripening, providing a consistent computational setting for systematic comparison of these competing mechanisms. 
Beyond these predefined barriers, our framework incorporates adaptive OTF barrier evaluation using machine-learned interatomic potentials (MLIPs). When previously unseen local coordination environments are encountered, the corresponding activation barriers are computed using a hierarchy of methods, ranging from frozen-geometry potential energy scans and Br{\o}nsted--Evans--Polanyi estimates\cite{van2009reactivity} to full nudged elastic band calculations\cite{henkelman2000improved}, and cached for reuse. 
The coexistence of predefined and OTF modes within a single codebase enables direct comparison between fixed and environment-aware kinetics.
The resulting MPI-parallel implementation \cite{Lubachevsky1988,Shim2005} provides a flexible and extensible platform for modeling GaN epitaxy, from atomistic processes to mesoscale morphology evolution.

\section{Computational Methodology and Workflow}

\begin{figure*}[t]
\centering
\includegraphics[width=0.95\textwidth]{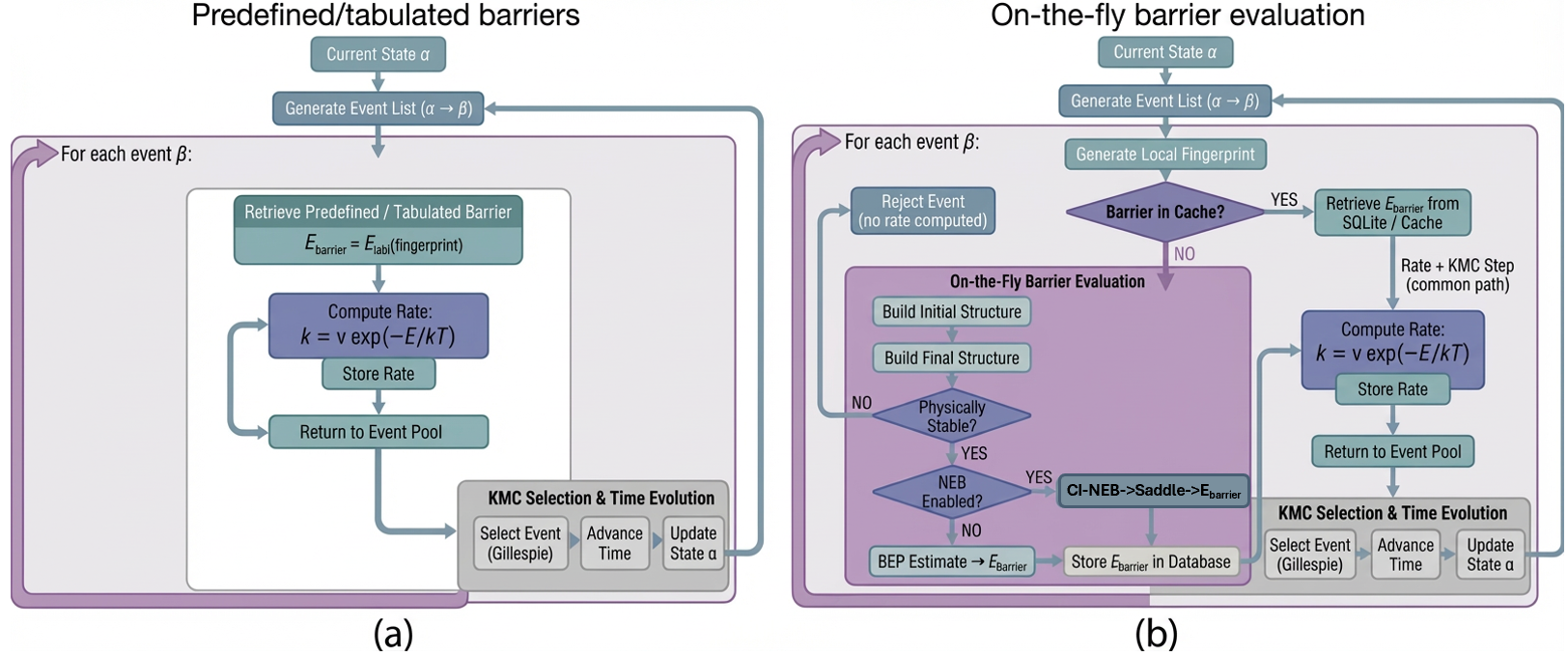}
\caption{Schematic comparison of kinetic Monte Carlo (KMC) workflows. (a) Conventional lattice KMC using predefined or tabulated activation barriers, where event rates are retrieved directly from a static database. (b) On-the-fly (OTF) KMC framework employed in this work, in which activation barriers are evaluated dynamically when a previously unseen local configuration is encountered. If the local fingerprint is not found in the stored cache, barriers are computed through either nudged elastic band (NEB), potential energy scan (PES), or Br{\o}nsted--Evans--Polanyi (BEP) estimates, after which the result is stored for subsequent reuse.}
\label{fig:workflow}
\end{figure*}

The KMC framework developed here follows the underlying lattice-based KMC algorithm as implemented by Chugh \textit{et al.}\cite{chugh2017lattice}, while incorporating an extension for flexible activation barrier evaluation. Specifically, the present implementation operates in two modes within a single codebase: (i)~predefined activation barriers drawn from a tabulated catalog, and (ii)~adaptive on-the-fly (OTF) barrier evaluation. This allows systematic studies with fixed barrier sets, as well as simulations in which barriers are computed dynamically as new configurations are encountered. 
In predefined-barrier mode, sketched in Fig.~\ref{fig:workflow}(a), activation energies are obtained from DFT calculations, drawn from the catalog reported in Ref.~\cite{chugh2017lattice}. This mode is used here to investigate ES barrier effects, desorption, temperature-dependent morphology evolution, and Ostwald ripening within a consistent, static framework. All simulations target homoepitaxial GaN(0001) growth on a flat substrate; the substrate--adlayer interaction is thus fully encoded in the DFT-derived barrier catalog. A vicinal or stepped substrate can be initialized by modifying the starting lattice occupancy, after which the existing step-edge kinetics apply without code modification.

In OTF mode, activation barriers are evaluated dynamically whenever a previously unseen local configuration is encountered, as shown in the flowchart in Fig.~\ref{fig:workflow}(b). At each KMC step, a list of possible events is generated from the current lattice configuration and each event is characterized using a discrete local atomic fingerprint. The fingerprint encodes the species and occupancies within a finite neighborhood centered on the diffusing atom, defined by two neighbor shells. The first shell (nn1) consists of the three tetrahedral bonding sites in the adjacent atomic layer, i.e., the nearest neighbors in the GaN wurtzite structure (as depicted in Fig.~\ref{fig:nn_shells}). The second shell (nn2) comprises the six same-sublattice lateral neighbors at the in-plane nearest-neighbor distance $a \approx 3.19$~\AA, capturing the local step-edge and island geometry. The fingerprint is species-dependent: for Ga hops, it records the number of N atoms and Ga adatoms (AdGa) in the bonding layer below (nn1) and the number of Ga atoms (nn2); for N hops, it records the number of Ga atoms in the bonding layer above (nn1) and lateral same-sublattice N neighbors (nn2). Events sharing the same discrete fingerprint are treated as equivalent and reuse the same activation barrier.

For each event, the fingerprint is used to query a multi-tier cache consisting of an in-memory dictionary and a persistent SQLite database. In predefined-barrier mode [Fig.~1(a)], the database simply acts as a lookup table. In OTF mode [Fig.~1(b)], if a matching entry is found, the activation energy is retrieved directly; otherwise, an on-the-fly barrier evaluation procedure is triggered. To compute new barriers, the initial and final atomic configurations are constructed and relaxed to ensure convergence to physically meaningful local minima. Configurations that yield unstable or unphysical geometries are discarded without assigning a rate. For new configurations, activation barriers can be evaluated using one of several methods: a nudged elastic band (NEB) calculation, in which intermediate images are relaxed while constrained to follow the reaction path; a frozen-geometry potential energy scan (PES) along the hop path; or a Brønsted--Evans--Polanyi (BEP) estimate, in which the activation barrier is expressed as $E_a = E_0 + \alpha \Delta E$. Here, $\Delta E$ is the reaction energy defined as the difference between the relaxed initial and final states, and $E_0$ and $\alpha$ are the reference barrier and transfer coefficient, respectively, which can be treated as linear fitting parameters. In our calculations, we set $E_0$ to the barrier value obtained from DFT for a representative near-symmetric event within the same nn1 class, and $\alpha = 0.5$,  corresponding to an approximately symmetric transition state. The newly computed barrier is then stored in the database and reused for all subsequent occurrences of the same fingerprint (equivalent events), ensuring computational efficiency over long simulations. Multiple MLIPs can be selected to perform these calculations, including pretrained universal models such as CHGNet\cite{deng2023chgnet}, used in this work, MACE\cite{batatia2022mace}, M3GNet\cite{chen2022universal}, and PFP\cite{PFP2022}, as well as custom-trained potentials (e.g., NequIP\cite{batzner20223}, ACE\cite{drautz2019atomic}, SNAP\cite{thompson2015spectral}). 

To ensure scalability, the code employs domain decomposition with MPI parallelization and halo synchronization\cite{Shim2005, Lubachevsky1988}. The root MPI rank performs barrier evaluations when required, while worker ranks advance the KMC simulation using cached activation energies. This separation maintains efficiency during high event-rate simulations and enables system sizes relevant to experimentally realistic GaN growth.

\section{Results and Discussion}

We employ our lattice-based KMC implementation to investigate the growth kinetics and resulting surface morphologies of GaN(0001) during the early stages of thin-film deposition. Simulations are performed on a $128 \times 128$ lattice with periodic boundary conditions in the lateral directions, allowing the development of extended surface features while minimizing finite-size effects. Unless otherwise stated, a total of $0.6~\mathrm{ML}$ is deposited at a flux of $0.2~\mathrm{ML~s^{-1}}$ (one monolayer (ML) corresponds to half of a completed GaN bilayer). Growth conditions are parameterized by the flux ratio, defined as the Ga flux divided by the total incoming flux; values exceeding 0.5 correspond to Ga-rich environments. We focus on a slightly Ga-rich regime with a flux ratio of 0.535, representative of experimentally accessible growth conditions.
All reported results have been obtained by averaging over up to 10 independent simulation trajectories.
The principal morphological features are consistent across the different runs, with only minor run-to-run variations, and the conclusions are unaffected by stochastic differences between trajectories.

This section is organized as follows. We first analyze growth kinetics using predefined (tabulated) activation barriers and discuss the main morphological phenomena, namely, triangular island formation, Ostwald ripening, ES barrier effects, and desorption-driven island dynamics. We then present results obtained with OTF barrier evaluation, and assess the impact of dynamically computed barriers on surface morphology.

\subsection{KMC Calculations Using Predefined Barriers}

\subsubsection{Emergence of Triangular-Shaped Islands}

\begin{figure}[t]
    \centering
    \includegraphics[width=1.0\linewidth]{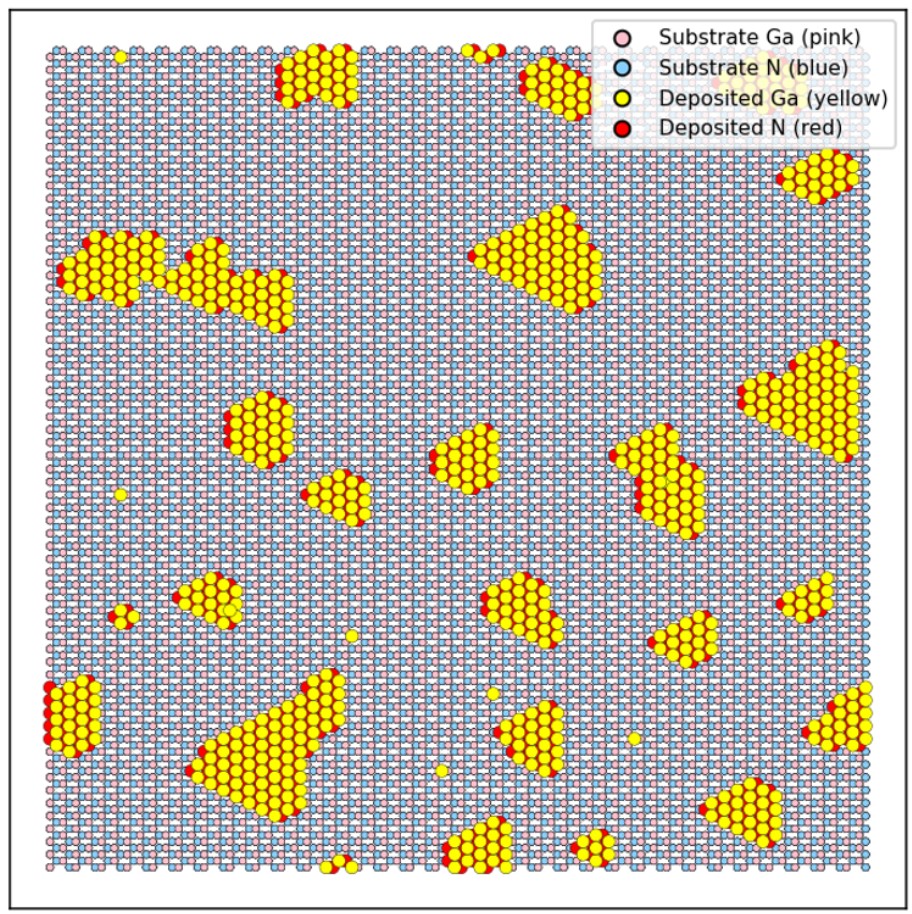}
    \caption{Final snapshot of the GaN(0001) surface morphology from KMC simulations during submonolayer growth on a $128 \times 128$ substrate at $700~\mathrm{K}$, with deposition flux $0.2~\mathrm{ML~s^{-1}}$ and flux ratio 0.535 (Ga-rich). Compact, triangular-shaped islands with a common crystallographic orientation are evident.}
    \label{fig:kmc_lit}
\end{figure}

Figure~\ref{fig:kmc_lit} shows the surface morphology at $0.6~\mathrm{ML}$ coverage and a substrate temperature of 700~K, obtained using predefined activation barriers. 
As deposition proceeds, isolated adatoms nucleate into islands that grow with increasing coverage, with the resulting surface characterized by compact, triangular-shaped islands with a uniform crystallographic orientation.
Similar island shapes have been reported in atomistic simulations of GaN growth \cite{chugh2017lattice, ranganathan2025ga} and are consistent with scanning tunneling microscopy observations of GaN(0001) surfaces grown under Ga-rich conditions\cite{xie1999anisotropic}, validating the present KMC framework. Diffusion energy anisotropy is encoded in the DFT-derived barrier catalog, which reflects the threefold symmetry of the GaN(0001) wurtzite surface; the triangular island morphology with uniform crystallographic orientation arise from anisotropic step-edge attachment and detachment kinetics, consistent with the lattice geometry.

\subsubsection{Ostwald Ripening}

When deposition is interrupted, island populations continue to evolve through thermally activated surface diffusion, leading to Ostwald ripening, in which mass is transferred from smaller, high-curvature islands to larger, energetically favored ones, reducing the total surface free energy. This process modifies the island size distribution and surface roughness and is particularly relevant to GaN MBE growth. In practical growth conditions, deposition commonly proceeds via shuttered flux sequences rather than continuous flux \cite{nicholls2023high}. During flux-off intervals, adatom diffusion and inter-island mass transport persist, enabling surface reorganization in the absence of additional material supply. Accurately modeling such conditions therefore requires a framework capable of capturing both deposition-driven kinetics and post-deposition relaxation within a unified description.

To investigate these effects, we perform KMC simulations to a coverage of 0.6~ML at 750~K, after which the external flux is terminated. The system is then allowed to evolve for $5$~s of simulated post-deposition time without further adatom injection. During this relaxation stage, no new material is supplied to the surface, and evolution is governed entirely by thermally activated diffusion, detachment, and reattachment processes.

\begin{figure}[t]
    \centering
    \includegraphics[width=1.0\linewidth]{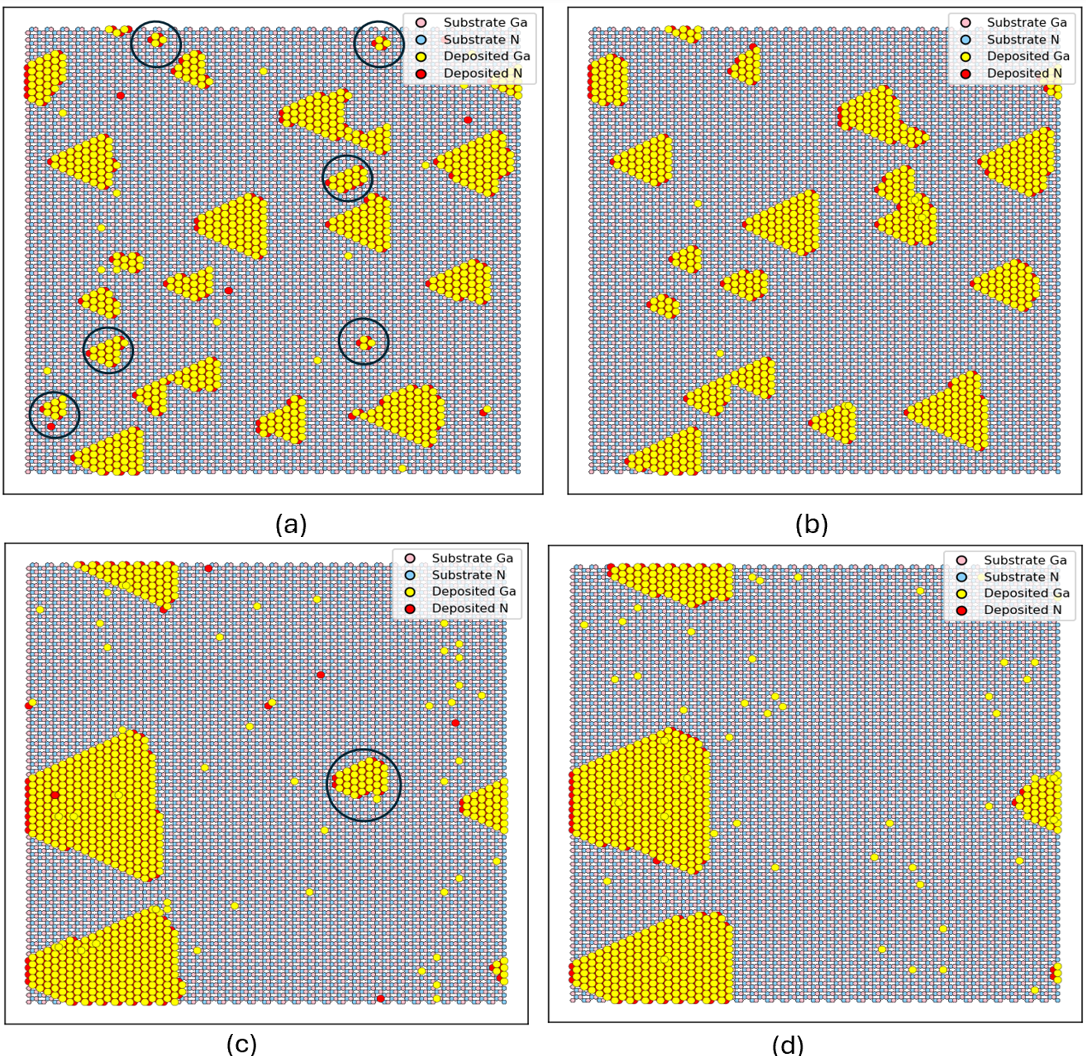}
    \caption{GaN(0001) surface morphology at the end of deposition and after 5~s of post-deposition evolution at two temperatures. (a)~End of active deposition at 750~K; compact triangular islands are visible. (c)~End of active deposition at 950~K; fewer, larger islands reflecting enhanced diffusivity. (b)~After 5~s of post-deposition evolution at 750~K: small islands [encircled in~(a)] have dissolved in favor of larger ones, illustrating Ostwald ripening. (d)~After 5~s of post-deposition evolution at 950~K: the island encircled in~(c) has dissolved into adjacent larger islands.}
    \label{fig:ostwald}
\end{figure}

As shown in Fig.~\ref{fig:ostwald}, a pronounced reduction in island density occurs during the post-deposition evolution. Several smaller islands present at the end of active growth [Fig.~\ref{fig:ostwald}(a)] progressively shrink and ultimately disappear, while larger islands grow through the incorporation of diffusing adatoms released from dissolving clusters. For clarity, the islands that vanish during relaxation are encircled in black in Fig.~\ref{fig:ostwald}(a) and are absent after the 5~s post-deposition interval in Fig.~\ref{fig:ostwald}(b). This behavior is characteristic of Ostwald ripening, 
resulting in a coarsened surface morphology and an increased average island size.

Repeating the simulation at $950~\mathrm{K}$ reveals how higher growth temperatures promote coarser, more well-defined island morphologies. As shown in Fig.~\ref{fig:ostwald}(c), the surface at the end of deposition features fewer and larger islands compared to the 750~K case [Fig.~\ref{fig:ostwald}(a)], consistent with the longer adatom diffusion length at elevated temperature. Figures~\ref{fig:ostwald}(c) and~(d) illustrate how a smaller island (encircled in black) dissolves in favor of the surrounding larger islands during the 5~s post-deposition evolution.

\subsubsection{Effect of Ehrlich-Schwoebel Barriers on Growth Morphology}

\begin{figure}[t]
    \centering
    \includegraphics[width=1.0\linewidth]{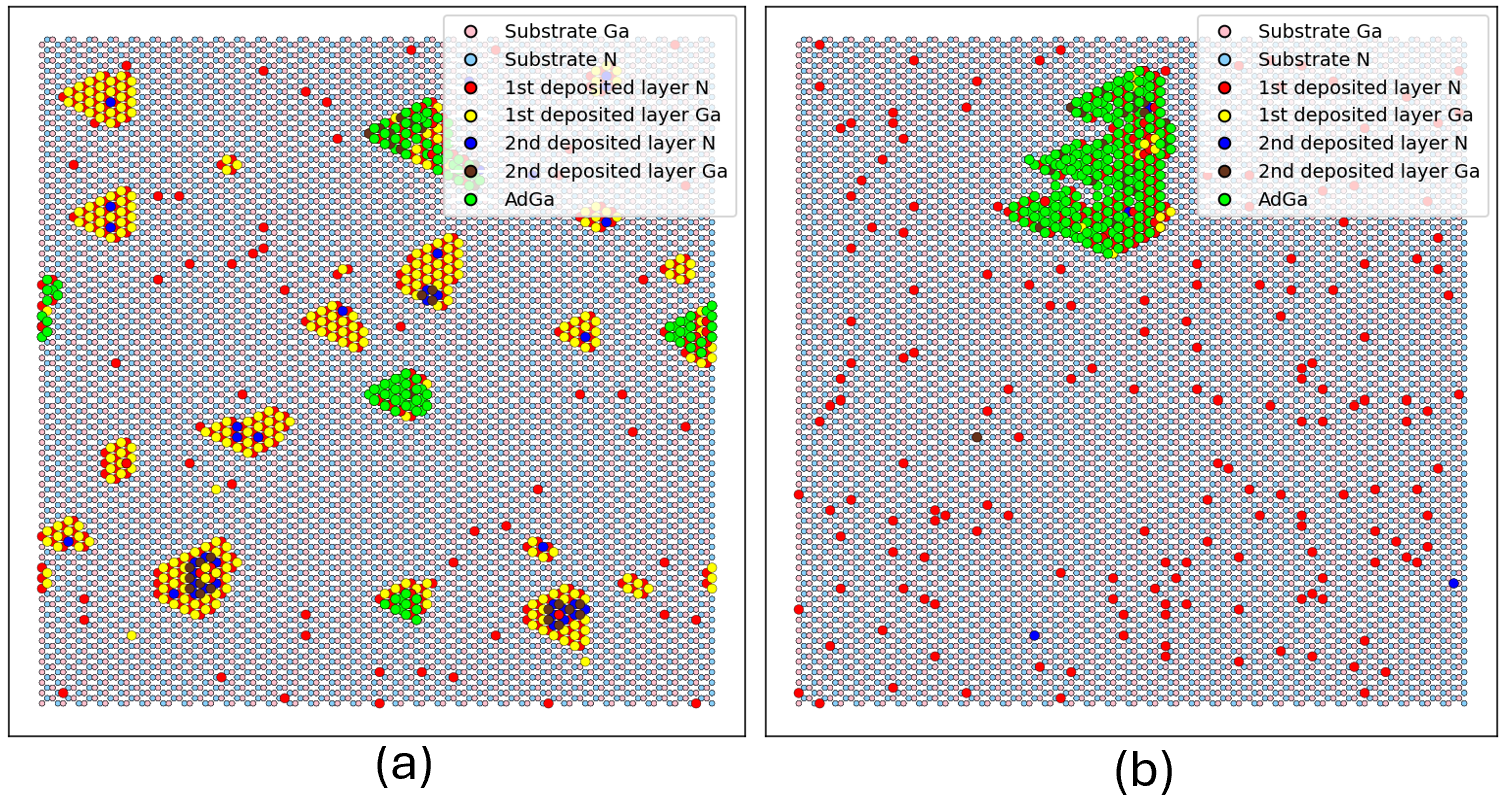}
    \caption{Surface morphologies obtained with the ES barrier activated. 
    (a)~$T = 750$~K, showing multiple compact islands and kinetically rough growth. 
    (b)~$T = 1000$~K, showing coarsening into a single irregular multilayer island.}
    \label{fig:SB}
\end{figure}

The activation of the ES barrier introduces an asymmetry in interlayer mass transport by inhibiting the downward diffusion of adatoms across step edges. This kinetic constraint generates an effective uphill mass current, promoting the accumulation of material on upper terraces. Such step-edge barriers are known to play a critical role in the morphological evolution of GaN(0001) surfaces, as discussed in Ref. \cite{kaufmann2016critical}. 

In the present simulations, we employ ES barrier energies of 
$E_{\mathrm{ES}}^{\mathrm{Ga}} = 0.15$~eV, 
$E_{\mathrm{ES}}^{\mathrm{AdGa}} = 0.15$~eV, 
and $E_{\mathrm{ES}}^{\mathrm{N}} = 0.25$~eV 
 to account for species-dependent interlayer diffusion asymmetry. These values represent conservative lower bounds relative to the previously reported values in the literature~\cite{zywietz1998adatom, turski2019unusual, kaufmann2016critical}, and are chosen to illustrate the strong morphological sensitivity to even modest interlayer diffusion asymmetry.
 The resulting surface morphologies at two representative temperatures are shown in Fig.~\ref{fig:SB}.

At $T = 750$~K, surface diffusion is active but remains significantly constrained by the ES barrier. The limited adatom diffusion length leads to frequent secondary nucleation events before existing islands can fully coalesce. As shown in Fig.~\ref{fig:SB}(a), the surface exhibits a high density of relatively compact islands distributed across the substrate. The inhibited step-edge crossing promotes local material accumulation and early-stage multilayer and AdGa layer formation, producing kinetically rough growth characterized by lateral island proliferation and incomplete coalescence.

At higher temperatures ($T = 1000$~K), diffusion becomes more effective, allowing adatoms to sample a larger surface area before incorporation. 
This suppresses secondary nucleation and promotes coarsening through preferential growth of larger islands. 
As observed in Fig.~\ref{fig:SB}(b), the morphology evolves toward a lower island density, dominated by a single large, irregularly shaped cluster. The reduced number of nuclei and the accumulation of material within the dominant island reflect diffusion-driven coarsening, while the ES barrier continues to bias attachment toward ascending steps, facilitating multilayer buildup within the growing structure. 

\subsubsection{Desorption Effects}

Desorption competes with deposition and diffusion during GaN(0001) MBE growth, influencing both the net growth rate and island morphology. Species-specific desorption activation barriers are implemented with coordination dependence: Ga adatoms ($E_{\mathrm{des}}^{\mathrm{Ga}} = 2.0$~eV base, $+0.5$~eV per N neighbor, $+0.2$~eV per AdGa), N atoms ($E_{\mathrm{des}}^{\mathrm{N}} = 2.5$~eV base, $+0.6$~eV per Ga), and AdGa ($E_{\mathrm{des}}^{\mathrm{AdGa}} = 1.5$~eV base, $+0.4$~eV per Ga), establishing the stability hierarchy N~$>$~Ga~$>$~AdGa.\cite{bruno2010adsorption, lymperakis2024desorption} 
Alongside these single-atom channels, a molecular N$_2$ desorption pathway is activated whenever two adjacent surface N atoms are both free of Ga coverage above them. The paired-N$_2$ barrier (1.5~eV) lies 1.0~eV below the isolated-N barrier, mimicking the experimentally observed dominance of associative desorption, whereby nitrogen leaves the surface primarily in molecular form\cite{karpov2000surface}. 

\begin{figure}[htbp]
    \centering
    \includegraphics[width=1.0\linewidth]{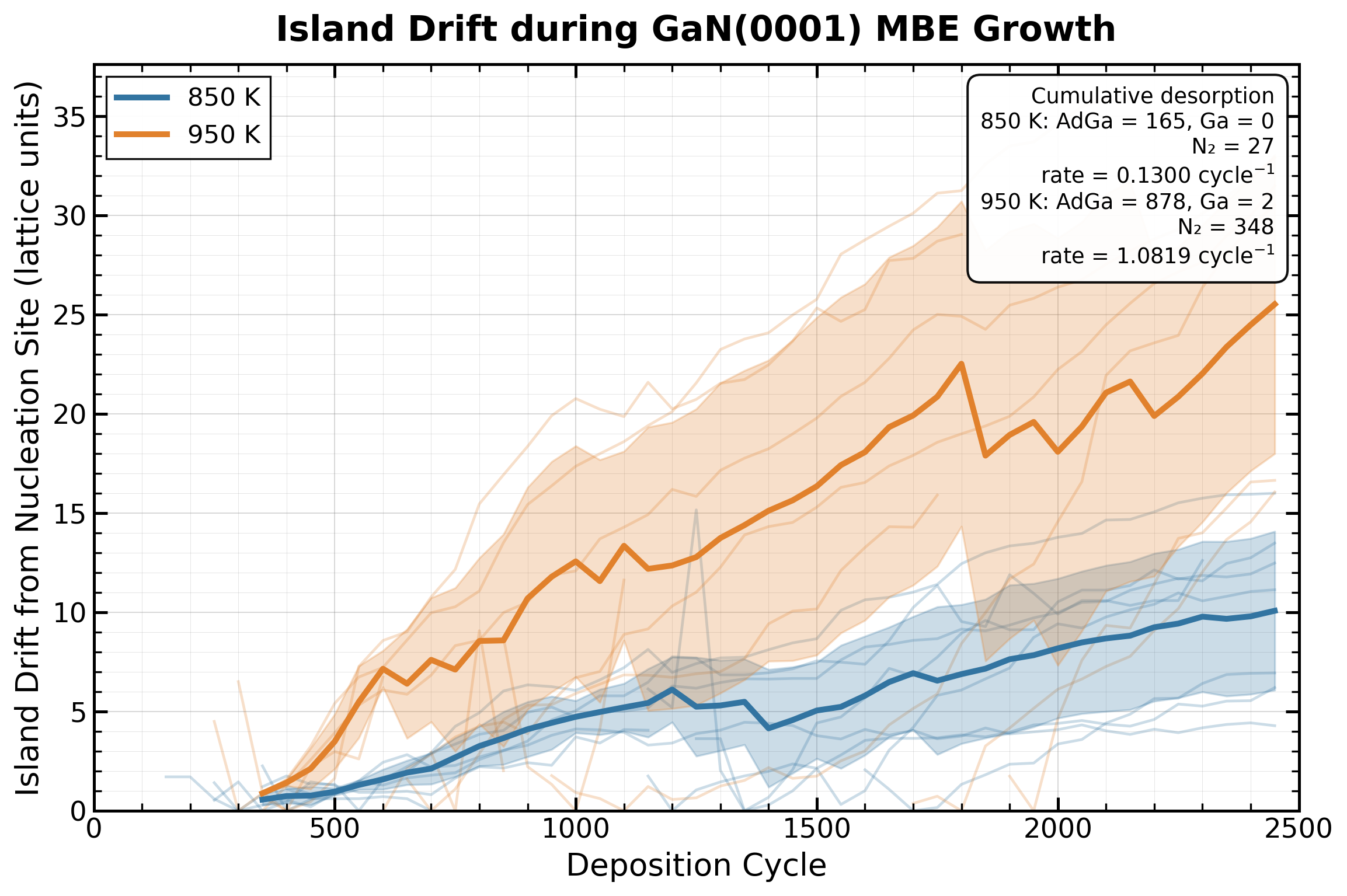}
    \caption{Mean island drift from nucleation site versus deposition cycle at 850~K (blue) and 950~K (orange). Islands are tracked from the snapshot at which the cluster first exceeds 15~atoms. Shaded bands show $\pm 1\sigma$ across all tracked islands and individual island trajectories appear as faint lines. The inset reports cumulative desorption totals over the full simulation (2450~cycles): at 850~K, AdGa~$=165$, Ga~$=0$, N$_2 = 27$ molecules, total rate $= 0.13$~cycle$^{-1}$; at 950~K, AdGa~$=878$, Ga~$=2$, N$_2 = 348$ molecules, total rate $= 1.08$~cycle$^{-1}$. The mean drifts at the end of each run are approximately 10 and 25~lattice units, respectively. The wide 950~K band reflects the stochastic sensitivity of desorption-driven island walking to local N$_2$ desorption heterogeneity across the surface.}
    \label{fig:drift}
\end{figure}

The simulated cumulative desorption counts (Fig.~\ref{fig:drift} inset) show that AdGa desorbs the most, with approximately five times as many events at 950~K as at 850~K. Ga evaporation remains negligible at both temperatures. All nitrogen loss occurs via the N$_2$ channel. At 850~K it is moderate, but is substantially larger at 950~K, where the total N removed represents a significant fraction of the deposited flux and constitutes a stoichiometric drain large enough to measurably suppress net surface coverage. The combined desorption rate increases more than eightfold between the two temperatures. This increase exceeds what AdGa removal alone would produce and signals the growing role of the N$_2$ channel.

The actual temperature dependence of the AdGa desorption rate is significantly below a naive Arrhenius estimate based on the energy barrier, because the number of events per cycle is governed by coupled processes: N--Ga exchange controls how AdGa is generated at step edges, while competing re-incorporation via terrace diffusion continuously depletes the elevated Ga pool.\cite{ranganathan2025ga} AdGa itself arises either by direct Ga-on-Ga deposition or, more importantly, by N--Ga exchange, in which a diffusing N displaces a surface Ga upward by one bilayer. Exchange-generated AdGa sits at the step edge with no overlying Ga coverage, giving it the minimum desorption barrier and directly coupling N surface diffusion to Ga mass loss.

The N$_2$ desorption rate behaves differently from AdGa: it does not exhibit a simple Arrhenius dependence on temperature but is intrinsically coverage-dependent, because N$_2$ formation requires two adjacent unbound N adatoms---an event whose probability grows with the instantaneous surface density of free N. At 850~K this process proceeds at a quasi-steady pace and contributes only a modest net coverage deficit. At 950~K the picture is qualitatively different: the rate is slow in the early growth stage but accelerates sharply in later cycles, once islands are large enough that freshly deposited N frequently lands near pre-existing unbound N on adjacent terraces. This feedback, where more free N finds more free N, amplifies the loss rate as the simulation progresses and produces a substantial difference between deposited and on-surface coverage that is essentially absent at 850~K. The result is a growth nonlinearity that cannot be captured by a simple Arrhenius model, as the effective N-loss rate is shaped by the evolving surface morphology.

\begin{figure}[htbp]
    \centering
    \includegraphics[width=1.0\linewidth]{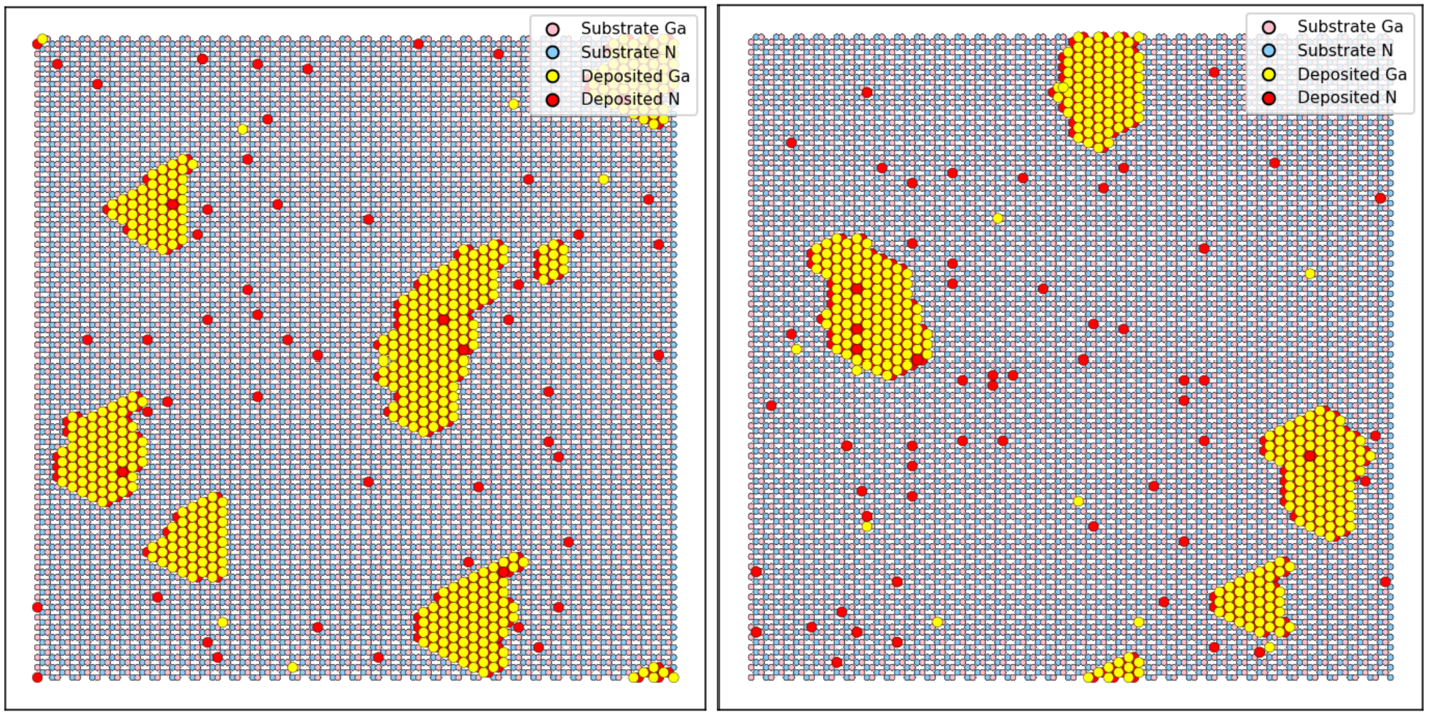}
    \caption{GaN(0001) surface morphology after 2450 deposition cycles with all desorption channels active. (a)~At 850~K, multiple islands of comparable size are present with roughened step edges, a signature of selective AdGa removal at kink and corner sites. N$_2$ desorption is moderate and does not significantly alter island density. (b)~At 950~K, a smaller number of coarser, asymmetrically shaped islands remain, reflecting the onset of desorption-driven island walking and the strongly accelerating N$_2$ loss from surrounding terraces in the late growth stage.}
    \label{fig:desorption}
\end{figure}

The surface morphology at 850~K, shown in Fig.~\ref{fig:desorption}(a), consists of several compact-to-irregular islands distributed across the substrate. Edge roughening is the main signature of desorption at this temperature: AdGa atoms at kink and corner sites have the lowest desorption barrier and are preferentially removed before they can stabilize protruding step segments, leading to ragged island outlines and suppressing the well-defined faceted triangles seen in desorption-free runs.

At 950~K [Fig.~\ref{fig:desorption}(b)], the morphology is substantially coarser: fewer, more irregularly shaped islands survive, with pronounced directional asymmetry in their outlines. This is the \emph{island walking} regime. N--Ga exchange at a trailing step edge elevates Ga to an AdGa position; the elevated Ga desorbs before it can diffuse back to a step, so the edge retreats by one site. Similarly, Ga adatoms reaching the geometrically stable leading vertex attach irreversibly. The asymmetry between irreversible growth at one edge and desorption-driven recession at the other displaces the island center of mass. The accelerating loss of N$_2$ from the surrounding terraces further disrupts the local stoichiometry of N needed to pin advancing Ga, strengthening the walking tendency rather than stabilizing compact shapes. The step-edge attachment asymmetry underlying this walking tendency has been directly confirmed by STM in GaN(0001) MBE: Xie~\textit{et al.}~\cite{xie1999anisotropic} demonstrate that type-A step edges (two dangling bonds per edge atom) accommodate adatoms more readily than type-B edges (one dangling bond). This produces anisotropic step-edge morphologies and triangular islands, confirming that crystallographically distinct step orientations interact differently with arriving adatoms. Such step-edge asymmetry provides the directional kinetic bias underlying island displacement.

In Fig.~\ref{fig:drift} we quantitatively track the island drift, making this contrast explicit. At 850~K, the islands accumulate a mean drift of about 10~lattice units from their nucleation sites, with a relatively wide spread across the island population. At 950~K the mean reaches approximately 25~lattice units, but the island-to-island variance is nearly as large as the mean itself. This heterogeneity arises because the walking rate depends sensitively on the local balance of N$_2$ desorption and N--Ga exchange, both of which fluctuate with the spatial distribution of unincorporated N across the terrace. Islands nucleating in locally N-depleted regions walk faster, while those surrounded by freshly arriving N walk slowly or barely move. Temperature sets the scale of the rates, but local stochastic chemistry ultimately determines which islands walk and by how much.

\subsection{KMC Calculations with On-the-Fly Barrier Evaluation}

\begin{figure}[htbp]
    \centering
    \includegraphics[width=\columnwidth]{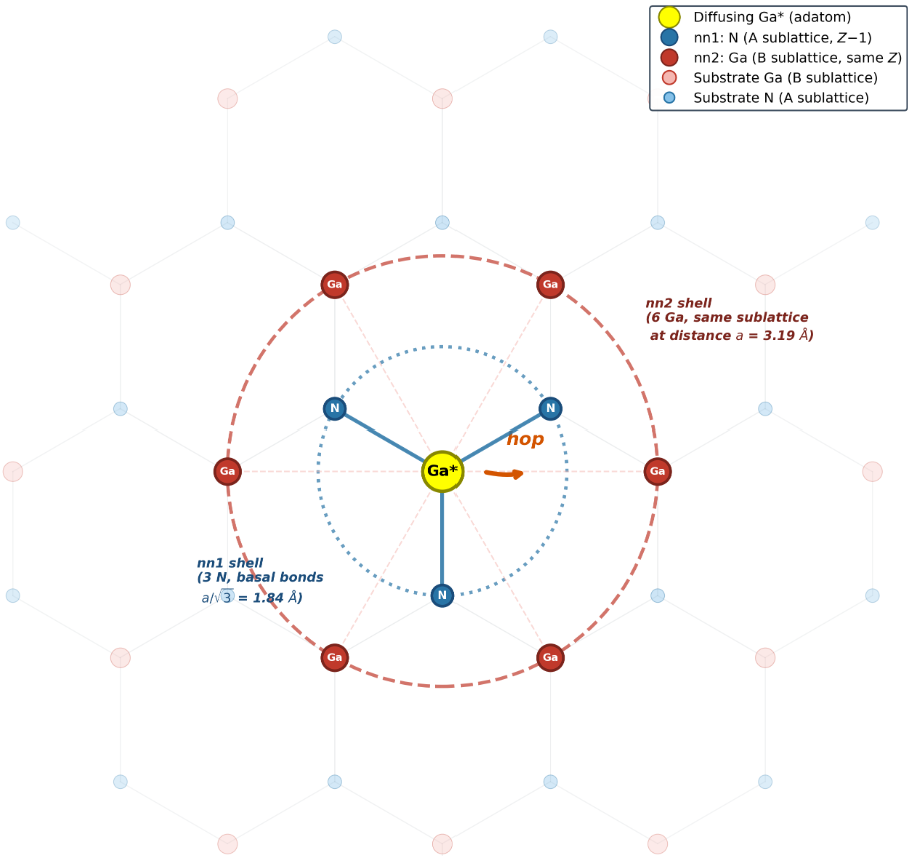}
    \caption{Schematic top view of the GaN(0001) wurtzite surface showing the nn1 and nn2 neighbor shells around a diffusing Ga adatom (Ga*, yellow). The nn1 shell (blue dotted circle) comprises 3~N atoms at the layer below ($Z{-}1$). The nn2 shell (red dashed circle) comprises 6~same-sublattice Ga atoms at distance~$a$ (${\approx}3.19$~\AA) in the same layer. The predefined catalog parameterizes barriers by nn1 only; OTF-KMC extends the descriptor to nn2, enabling distinct lateral environments to be resolved.}
    \label{fig:nn_shells}
\end{figure}

\begin{figure*}[t]
    \centering
    \includegraphics[width=0.95\textwidth]{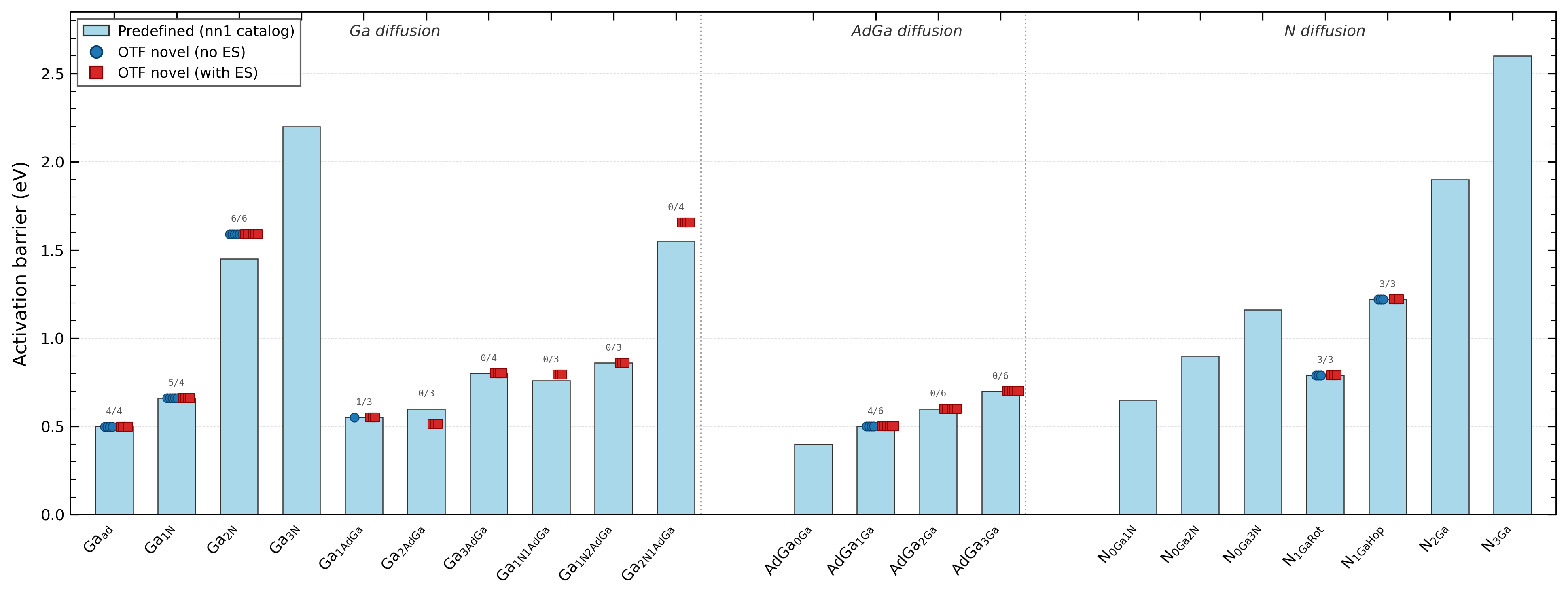}
    \caption{Complete predefined barrier catalog (21~events, light-blue bars) taken from the DFT-based tabulation of Chugh~\textit{et al.}\cite{chugh2017lattice}, grouped by diffusing species. Scatter points show novel configurations whose barriers were evaluated on-the-fly using the BEP relation\cite{deng2023chgnet}: blue circles correspond to simulations without ES barriers, red squares to simulations with ES barriers included. 
    The scatter points represent novel configurations (nn2$_\mathrm{Ga}{>}0$); their coincidence with predefined barriers for most event types is a physical consequence of near-zero BEP reaction energies ($\Delta E \approx 0$) for symmetric hop geometries, which maps the novel barrier back to $E_0$ regardless of nn2 coordination. Numbers above each bar indicate the count of novel configurations discovered (no~ES\,/\,with~ES). Significant deviations at Ga$_{2\mathrm{N}}$ ($+0.14$~eV), Ga$_{2\mathrm{N},1\mathrm{AdGa}}$ ($+0.11$~eV), and Ga$_{2\mathrm{AdGa}}$ ($-0.09$~eV) identify asymmetric step-edge environments where nonzero $\Delta E$ produces a measurably distinct barrier.}
    \label{fig:otf_barriers}
\end{figure*}

The predefined barrier catalog used in the previous sections, shown in Fig.~\ref{fig:otf_barriers}, parameterizes each diffusion event based solely on first-nearest-neighbor (nn1) coordination. For Ga diffusion, Ga$_{k\mathrm{N}}$ denotes a hop in which $k$~nitrogen atoms occupy the three tetrahedral bonding sites in the layer immediately below the diffusing Ga adatom---its true nearest-neighbors, forming short Ga--N bonds ($\approx 1.95$~\AA). 
The case Ga$_{\mathrm{ad}}$ ($k{=}0$) corresponds to an isolated adatom above the bare Ga surface. Mixed-species variants, Ga$_{k\mathrm{N},m\mathrm{AdGa}}$, additionally include $m$~adatoms (AdGa) occupying these nn1 bonding sites. For N diffusion, N$_{k\mathrm{Ga}}$ counts the $k$~Ga atoms at the three tetrahedral bonding sites in the layer immediately above the diffusing N, distinguishing rotation and hop channels at $k{=}1$. For AdGa diffusion, AdGa$_{k\mathrm{Ga}}$ counts $k$~Ga atoms in the next-bilayer positions immediately above the AdGa atom, representing coverage from the film growing on top. In all three cases, the counted atoms are the true nearest-neighbor bond partners in the GaN wurtzite structure; lateral same-sublattice Ga atoms at the nn2 distance ($\approx 3.19$~\AA) are not bonded to the diffusing species and are therefore excluded from the nn1 descriptor. This yields tabulated barriers that fully determine diffusion kinetics on flat terraces. However, all events sharing the same nn1 class are assigned an identical barrier irrespective of the lateral (same-sublattice) environment, meaning that terrace, step-edge, and kink geometries are treated equivalently. 

To resolve this limitation, we introduce an on-the-fly KMC capability that extends the local descriptor to include the second-nearest-neighbor (nn2) coordination shell, i.e., the six same-sublattice Ga sites at distance~$a$ forming a hexagonal ring around the diffusing atom (Fig.~\ref{fig:nn_shells}). When the simulation encounters an nn1$\times$nn2 combination that is absent from the predefined catalog, the activation barrier is computed dynamically as described below. Extending the descriptor from nn1 to nn1$\times$nn2 
expands the theoretical configuration space from 20 to 140 fingerprints without ES barriers, or from 29 to 497 with ES barriers active. Hence, an exhaustive pre-computation of all possible nn1$\times$nn2 barriers via nudged-elastic-band calculations would require up to 140 (without ES barriers) or 497 (with ES barriers) separate calculations per growth condition, most of which would never be encountered in a given simulation trajectory (see Section~S2 of the Supporting Information for more detail). The OTF-KMC approach resolves this by computing barriers on demand when novel configurations arise and caching them for reuse, making environment-aware simulations feasible.
Although the OTF framework is capable of computing activation barriers using full NEB calculations, in the present work we employ a Br{\o}nsted--Evansf--Polanyi (BEP) evaluation strategy in conjunction with a machine-learned interatomic potential, CHGNet\cite{deng2023chgnet}. This choice balances computational efficiency and physical consistency: the BEP relation described above retains the DFT-calibrated references barriers $E_0$ from the predefined catalog while introducing corrections associated with the  changes (i.e. $\Delta E$) introduced by extended local environments (nn2 configurations), without requiring explicit saddle-point optimization.

A BEP sensitivity analysis varying $\alpha$ from 0.25 to 0.75 shows that the three asymmetric OTF-discovered step-edge configurations (Ga$_{2\mathrm{N}}$, Ga$_{2\mathrm{AdGa}}$, Ga$_{2\mathrm{N},1\mathrm{AdGa}}$) exhibit total barrier variations of 0.086--0.14~eV across this interval (see Table~\ref{tab:bep_si} in the Supporting Information). Here we adopt the symmetric value $\alpha = 0.5$ as the standard reference for a near-symmetric transition state, consistent with the hop geometry of the GaN(0001) surface.

We denote these OTF-discovered environments as \emph{novel configurations}. Their naming follows the predefined convention, with the number of lateral Ga atoms in the nn2 shell included as an additional label. For example, Ga$_{2\mathrm{N}}$\,(nn2$_{\mathrm{Ga}}{=}j$) denotes a Ga hop with 2~N atoms below (identical nn1 coordination to the predefined Ga$_{2\mathrm{N}}$ entry), but with $j$~lateral same-sublattice Ga atoms in the nn2 shell. Because this shell captures the local step-edge geometry, each value of $j$ corresponds to a distinct bonding environment, from an isolated terrace ($j{=}0$, the predefined case) to a fully coordinated kink ($j{=}6$). As a concrete example, the predefined Ga$_{2\mathrm{N}}$ barrier is $1.45$~eV for a flat terrace, whereas the novel Ga$_{2\mathrm{N}}$\,(nn2$_{\mathrm{Ga}}{=}2$) configuration, representing a step-edge site with two lateral Ga neighbors, yields $1.59$~eV. This $0.14$~eV increase reflects the additional stabilization associated with the lateral Ga--Ga bonding, which makes detachment less favorable. This correction is absent in the nn1-only catalog, which would assign the same $1.45$~eV barrier to both environments. 

After removing symmetry-equivalent structures, the OTF-KMC simulation without ES barriers discovers 26~unique novel configurations across 7~event types, while the simulation including ES barriers yields 58 configurations across 14~types---a ${\sim}2.2{\times}$ increase consistent with the greater diversity of step-edge geometries generated by kinetic roughening. Figure~\ref{fig:otf_barriers} compares the 21~predefined barriers with the corresponding novel OTF configurations overlaid. For most events, the novel barriers cluster within ${\pm}0.01$~eV of their predefined counterparts, confirming the accuracy of the nn1 catalog on flat terraces. Significant deviations occur for Ga$_{2\mathrm{N}}$ ($+0.14$~eV), Ga$_{2\mathrm{N},1\mathrm{AdGa}}$ ($+0.11$~eV), and Ga$_{2\mathrm{AdGa}}$ ($-0.09$~eV), identifying the specific nn1 classes where lateral coordination produces a physically distinct barrier. In the presence of ES barriers, additional event classes are activated, including AdGa$_{2\mathrm{Ga}}$ and AdGa$_{3\mathrm{Ga}}$, as well as mixed-species Ga environments (Ga$_{1\mathrm{N},1\mathrm{AdGa}}$, Ga$_{3\mathrm{AdGa}}$) that are not encountered without ES barriers. This reflects the Ga-rich step-edge decoration that emerges when downward transport is inhibited.

\begin{figure}[htbp]
    \centering
    \includegraphics[width=1.00\linewidth]{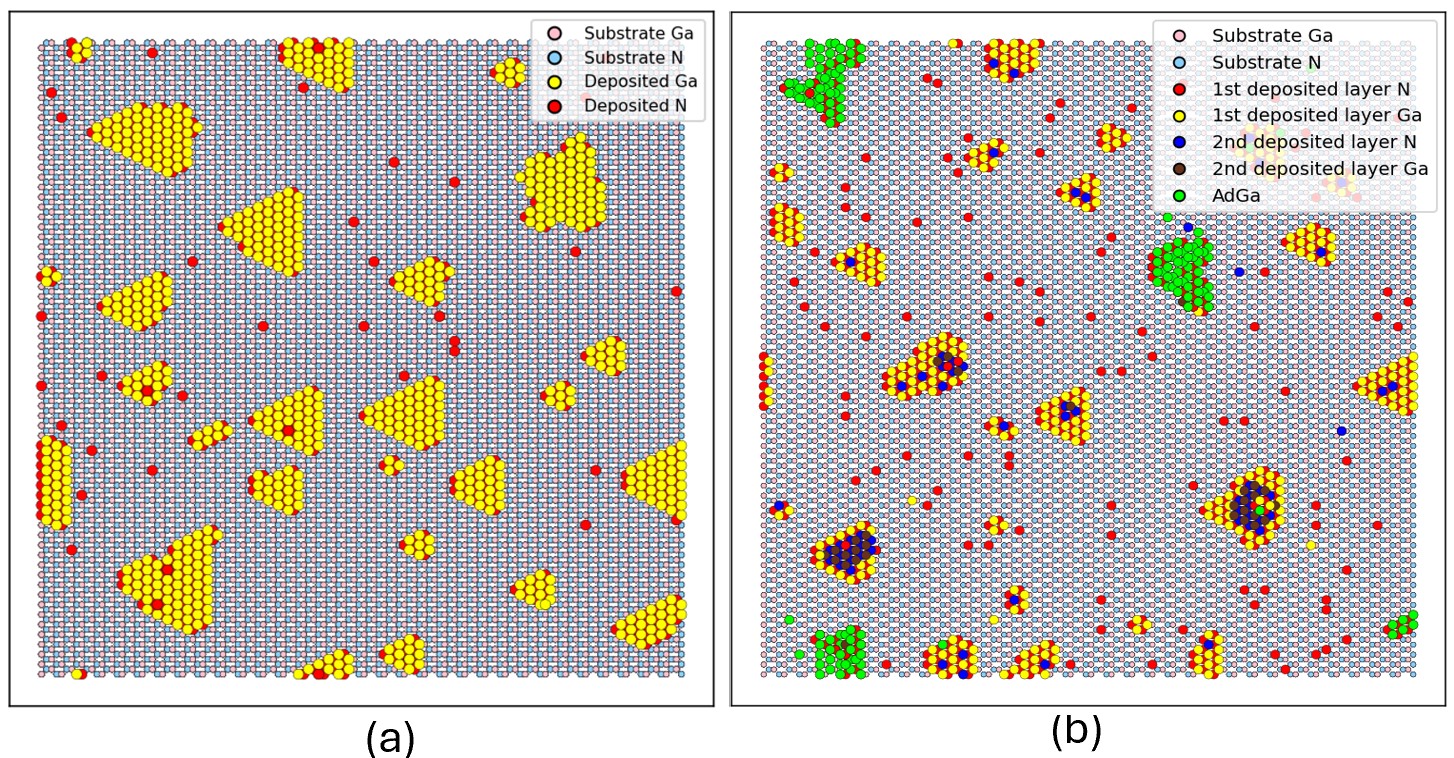}
    \caption{OTF-KMC surface morphologies of GaN(0001) at $0.6$~ML. (a)~Without ES barriers ($700$~K): compact first-layer Ga--N islands (yellow: 1st-layer Ga, red: N adatoms), preserving the same growth mode as the predefined-barrier result (Fig.~\ref{fig:kmc_lit}), but with fewer and noticeably larger islands with more regular, elongated perimeters. This behavior can be attributed to the $0.14$~eV increase in the step-edge barrier for Ga$_{2\mathrm{N}}$ novel configurations, which kinetically traps step-edge atoms and suppresses island-shape relaxation. (b)~With ES barriers ($750$~K): kinetic roughening, second-layer nucleation (blue: 2nd-layer N, dark: 2nd-layer Ga), and pronounced AdGa accumulation (green) at step edges and terrace boundaries. These features are consistent with Fig.~\ref{fig:SB}(a), but with larger AdGa patches reflecting the additional AdGa-decorated edge environments resolved by OTF.}
    \label{fig:otf_morphology}
\end{figure}

The morphological impact of OTF barriers is assessed by comparing the final growth surfaces with the predefined-barrier results of Figs.~\ref{fig:kmc_lit} and~\ref{fig:SB}. Figure~\ref{fig:otf_morphology} shows the OTF-KMC morphologies at $0.6$~ML coverage. Without ES barriers [Fig.~\ref{fig:otf_morphology}(a)], the OTF result preserves the same first-layer growth mode as the predefined-barrier result (Fig.~\ref{fig:kmc_lit}): both show compact Ga--N islands with scattered N adatoms, confirming that the OTF method correctly reproduces the terrace diffusion and island-attachment kinetics of the tabulated catalog. The key morphological difference is that the OTF run produces marginally fewer but larger islands, with more regular and elongated perimeters. This is physically consistent with the $0.14$~eV step-edge barrier increase computed for Ga$_{2\mathrm{N}}$ novel configurations: the larger barrier stabilizes Ga atoms at step edges, suppressing the detachment--reattachment dynamics that normally drive island rounding and coarsening. Step-edge atoms are kinetically trapped, inhibiting shape relaxation and yielding larger, more elongated island morphologies relative to those obtained with the predefined flat-terrace value of $1.45$~eV. With ES barriers activated [Fig.~\ref{fig:otf_morphology}(b)], the OTF simulation reproduces the kinetic-roughening regime of Fig.~\ref{fig:SB}(a): both show multilayer islands with pronounced second-layer nucleation and Ga-rich step-edge decoration. A quantitative difference is the more extended AdGa accumulations [highlighted in green in Fig.~\ref{fig:otf_morphology}(b)] in the OTF case, particularly visible as dense patches at step edges and terrace boundaries. These arise from the additional AdGa-decorated edge environments resolved by OTF (e.g., Ga$_{1\mathrm{N},1\mathrm{AdGa}}$, Ga$_{2\mathrm{AdGa}}$), which carry modified barriers relative to the predefined values and thus affect the local balance of AdGa attachment and detachment at step edges. Overall, the close agreement between predefined and OTF morphologies in both regimes supports the consistency of the OTF barrier evaluation, while the remaining quantitative differences at step edges reflect the physically meaningful corrections introduced by the novel configurations discovered.

\section*{Conclusions}

We have developed a lattice-based KMC framework for modeling GaN(0001) growth under molecular beam epitaxy conditions. The framework operates in two modes: (i)~a predefined activation barrier mode based on coordination-classified event catalogs and (ii)~an OTF mode in which activation barriers are evaluated dynamically according to the instantaneous local atomic environment, enabling both computational efficiency and adaptive treatment of complex, evolving surface configurations.

Using predefined barriers, the simulations reproduce key morphological trends observed in GaN growth\cite{chugh2017lattice}, including compact triangular island formation under diffusion-dominated conditions, Ostwald ripening during growth interruptions, and Ehrlich-Schwoebel barrier-induced multilayer nucleation. Incorporating desorption reveals the sensitivity of the morphology to coordination-dependent removal kinetics, leading to edge roughening at moderate temperatures and directed island translation at elevated temperatures via the N--Ga exchange mechanism.

The ability to resolve post-deposition evolution is particularly relevant for modeling realistic MBE shutter sequences, where alternating flux-on and flux-off intervals strongly influence nanoscale morphology\cite{nicholls2023high}. By explicitly simulating growth interruptions and subsequent ripening, the present framework connects atomistic kinetics with the pulsed and modulated growth protocols employed experimentally.

While predefined barrier catalogs offer transparency and computational efficiency, their reliance on fixed coordination classes limits the range of accessible environments. In particular, they cannot distinguish between terrace, step-edge, and kink geometries that share the same nn1 coordination. The OTF implementation addresses this limitation by dynamically evaluating activation barriers as new configurations arise, extending the model to complex surface evolution near step edges, island boundaries, and multilayer structures.

The framework developed here brings together coordination-dependent kinetics, step-edge effects, desorption, Ostwald ripening, and adaptive barrier evaluation within a lattice KMC scheme, linking atomistic energetics to mesoscale morphology in non-equilibrium GaN growth, and is readily extendable to other compound semiconductor systems. The framework is well-positioned to guide experimental optimization of growth windows, including temperature, III/V flux ratio, and epitaxial beam control, for controlling the transition between island, step-flow, and roughening growth modes relevant to AlGaN-based superlattice LED structures. The adaptive OTF capability also enables exploration of growth conditions for which predefined catalogs are insufficient, with predicted morphologies directly testable by \textit{in situ} STM or post-growth AFM.

\vspace{0.5em}  

\section*{Acknowledgments}

S.A. and C.V. acknowledge financial support from the Australian Research Council Centre of Excellence in Quantum Biotechnology (QUBIC) under grant number CE230100021, as well as support from the Australian Research Council under grant number DE220101147. Computational resources were provided by the National Computational Infrastructure and the Pawsey Supercomputing Research Centre through the National Computational Merit Allocation Scheme.

\section*{Data Availability Statement}

The datasets generated and analyzed during the current study, together with the KMC code developed for this work, are available from the corresponding author upon reasonable request. The codebase will be made publicly available as open-source software following the publication of this study.

\section{References}
\bibliographystyle{apsrev4-2}
\bibliography{references}
\clearpage
\setcounter{equation}{0}
\setcounter{table}{0}
\renewcommand{\theequation}{S\arabic{equation}}
\renewcommand{\thetable}{S\arabic{table}}
 
\section*{Supporting Information}

\subsection*{S1. BEP Transfer-Coefficient Sensitivity Analysis}

In the BEP relation $E_a = E_0 + \alpha\,\Delta E,$  $E_0$ is taken from the DFT/NEB predefined catalog and $\Delta E$ is the reaction energy defined as
the difference between the relaxed initial and final states obtained using CHGNet.
Table~\ref{tab:bep_si} reports $E_{\mathrm{BEP}}(\alpha)$ for the three OTF-discovered configurations with non-zero reaction energy
(Ga$_{2\mathrm{N}}$, Ga$_{2\mathrm{AdGa}}$, and Ga$_{2\mathrm{N},1\mathrm{AdGa}}$) evaluated at $\alpha = 0.25$, 0.50, and~0.75.
These asymmetric step-edge configurations are sensitive to the choice of transfer coefficient:
the predicted barriers exhibit total variations of 0.08--0.14~eV as $\alpha$ is varied.
We adopt the symmetric value $\alpha = 0.5$, the standard Marcus-like choice for a near-symmetric transition state,
consistent with the hop geometry of the GaN(0001) surface and appropriate in the absence of explicit transition-state asymmetry information.

\begin{table}[ht]
\caption{BEP transfer-coefficient sensitivity for the three OTF-discovered
configurations with non-zero reaction energy. $E_0$: DFT/NEB reference barrier;
$\Delta E$: CHGNet reaction energy; $E_{\rm BEP}(\alpha)$: predicted barrier;
$\Delta$: total spread across $\alpha \in [0.25,\,0.75]$.}
\label{tab:bep_si}
\begin{ruledtabular}
\begin{tabular}{lcccccc}
Event & $E_0$ & $\Delta E$ & $\alpha{=}0.25$ & $\alpha{=}0.50$ & $\alpha{=}0.75$ & $\Delta$ \\
      & (eV)  & (eV)       & (eV) & (eV) & (eV) & (eV) \\
\hline
Ga$_{2N}$               & 1.450 & $+$0.280 & 1.520 & 1.590 & 1.660 & 0.140 \\
Ga$_{2\rm AdGa}$        & 0.600 & $-$0.172 & 0.557 & 0.514 & 0.471 & 0.086 \\
Ga$_{2N,1\rm AdGa}$    & 1.550 & $+$0.213 & 1.603 & 1.657 & 1.710 & 0.107 \\
\end{tabular}
\end{ruledtabular}
\end{table}

\subsection*{S2. Configuration-Space Analysis of the KMC Fingerprint}
 
Each kinetic Monte Carlo (KMC) diffusion event is characterised by a count-based
local-environment fingerprint recording the number of each atomic species within
crystallographic shells centred on the diffusing atom.
Three shells are defined by the GaN wurtzite (0001) geometry: the \textbf{nn1}
tetrahedral bond shell (3~sites in the adjacent layer at $r_1\approx1.95$~\AA),
the \textbf{nn2} same-sublattice lateral first ring (6~sites at
$a\approx3.19$~\AA), and the \textbf{nn3} lateral second ring
(6~sites at $\sqrt{3}\,a\approx5.52$~\AA).
The species considered are Ga (gallium), N (nitrogen), and AdGa (adsorbed Ga
adatom).
 
The number of distinct count-based fingerprints for a shell of $L$~sites with
$n_s$ distinguishable occupancy types (including the vacant state) follows from
the stars-and-bars identity:
\begin{equation}
    \Omega(L,\, n_s) = \binom{L + n_s - 1}{n_s - 1}.
    \label{eq:omega}
\end{equation}
For each species: the Ga \textbf{nn1} shell
($L{=}3$, occupants $\{\varnothing,\mathrm{N},\mathrm{AdGa}\}$, $n_s{=}3$)
gives $\Omega{=}\binom{5}{2}{=}10$;
N and AdGa \textbf{nn1} shells ($L{=}3$, $n_s{=}2$) give $\Omega{=}4$ each.
Lateral \textbf{nn2}/\textbf{nn3} shells with binary occupancy give $\Omega{=}7$
per shell; with ternary occupancy $\{\varnothing,\mathrm{Ga},\mathrm{AdGa}\}$
when Ehrlich-Schwoebel (ES) barriers are active, $\Omega{=}28$.
 
For N diffusion, environments with total \textbf{nn1} coordination
$k_{\mathrm{tot}}\in\{1,2\}$ (where $k_{\mathrm{tot}}\in\{0,1,2,3\}$ counts
the number of occupied sites in the diffusing N atom's 3-site \textbf{nn1}
shell) support both translational hop and in-plane rotation events (distinct
barriers). $k_{\mathrm{tot}}\in\{0,3\}$ support hops only, yielding 6~distinct
N event subtypes without ES barrier, or 15 with ES barrier (as AdGa may also occupy the
3~\textbf{nn1} Ga sites, expanding the environment set from 4 to 10).
 
The resulting totals are given in Tables~\ref{tab:configs_nosb} and
\ref{tab:configs_sb}.
The 20~fingerprints at the \textbf{nn1} level without ES barrier map to the 21~barriers
in the predefined catalog employed in this work: the N$_{1\mathrm{Ga}}$
fingerprint supports two distinct event types (rotation and translational hop)
with different activation energies (0.79~eV and 1.22~eV respectively), yielding
21~total barriers from 20~fingerprints.
Extending the descriptor to \textbf{nn2} raises the theoretical maximum to 140
without ES barrier and 497 with the barrier, factors of $7\times$ and $24\times$, respectively, over the predefined catalog.

These values represent theoretical upper bounds.
At a coverage of 0.6~ML (monolayer) and the simulated temperatures
(850--950~K), the mean lateral site occupancy
$\bar{\rho}\approx0.3$--$0.4$ yields an average $\langle j\rangle\approx1.8$--$2.4$ for
the \textbf{nn2} count, strongly suppressing high-coordination configurations
($j\geq4$).
The OTF-KMC simulations in the main text (using the \textbf{nn1}$\times$\textbf{nn2}
descriptor) encountered 26~novel fingerprints out of a maximum of 140 without ES barrier
($19\%$) and 58~out of 497 with this barrier ($12\%$).
Crucially, the specific fingerprints encountered cannot be determined
\emph{a~priori}, as they depend on the stochastic growth trajectory.

Exhaustive pre-computation of all possible \textbf{nn1}$\times$\textbf{nn2}
barriers via NEB would require up to $\sim\!500$ calculations for a given
local-environment definition, most of which would not be sampled in a typical
KMC trajectory. While such a one-off pre-computation may be feasible for a
fixed descriptor at this level of complexity, it becomes increasingly inefficient and difficult to generalise as the configuration space is extended. In particular, inclusion of additional coordination shells or surface topology (e.g., step edges, kinks, or rough morphologies) leads to a combinatorial increase in the number of distinct local environments, rendering exhaustive pre-computation impractical. In contrast, OTF evaluation computes barriers on demand and caches them for reuse, providing a scalable and flexible approach for environment-aware GaN growth simulations.

\begin{table}[ht]
\caption{Maximum distinct event fingerprints without ES barriers (Ga, N,
AdGa events; all lateral shells binary). N counts include hop and rotation
subtypes. The 20 \textbf{nn1} fingerprints correspond to the 21-barrier
predefined catalog.}
\label{tab:configs_nosb}
\begin{ruledtabular}
\begin{tabular}{lccc}
 Species & \textbf{nn1} only & \textbf{nn1}$\times$\textbf{nn2}
         & \textbf{nn1}$\times$\textbf{nn2}$\times$\textbf{nn3} \\
\midrule
 Ga   & 10 & $10\times 7 = 70$  & $10\times 7\times 7 = 490$  \\
 N    & 6  & $6\times 7  = 42$  & $6\times 7\times 7  = 294$  \\
 AdGa & 4  & $4\times 7  = 28$  & $4\times 7\times 7  = 196$  \\
\midrule
 Total & \textbf{20} & \textbf{140} & \textbf{980} \\
\end{tabular}
\end{ruledtabular}
\end{table}
 
\begin{table}[ht]
\caption{Maximum distinct event fingerprints with ES barriers active.
Ga and AdGa lateral shells expand to ternary occupancy ($\Omega{=}28$ per
shell); N event subtypes increase from 6 to 15.}
\label{tab:configs_sb}
\begin{ruledtabular}
\begin{tabular}{lccc}
 Species & \textbf{nn1} only & \textbf{nn1}$\times$\textbf{nn2}
         & \textbf{nn1}$\times$\textbf{nn2}$\times$\textbf{nn3} \\
\midrule
 Ga   & 10 & $10\times 28 = 280$  & $10\times 28\times 28 = 7{,}840$  \\
 N    & 15 & $15\times 7  = 105$  & $15\times 7\times 7  = 735$       \\
 AdGa &  4 & $4\times 28  = 112$  & $4\times 28\times 28 = 3{,}136$   \\
\midrule
 Total & \textbf{29} & \textbf{497} & \textbf{11{,}711} \\
\end{tabular}
\end{ruledtabular}
\end{table}

\end{document}